\documentclass[conference]{IEEEtran}
\IEEEoverridecommandlockouts
\usepackage{cite}
\usepackage{amsmath,amssymb,amsfonts}
\usepackage{algorithmic}
\usepackage{graphicx}
\usepackage{textcomp}
\usepackage{xcolor}
\usepackage{orcidlink}
\usepackage{multirow}
\usepackage{tikz} 
\usepackage{booktabs}
\usepackage{caption}
\usetikzlibrary{positioning}
\newcommand{\xb}{\ensuremath{\boldsymbol{x}}}
\newcommand{\yb}{\ensuremath{\boldsymbol{y}}}
\newcommand{\nb}{\ensuremath{\boldsymbol{n}}}
\newcommand{\rb}{\ensuremath{\boldsymbol{r}}}

\newcommand{\Nb}{\ensuremath{\boldsymbol{\mathsf{N}}}}

\newcommand{\Phib}{\ensuremath{\boldsymbol{\Phi}}}
\newcommand{\thetab}{\ensuremath{\boldsymbol{\theta}}}

\newcommand{\eC}{\mathbb{C}}

\newcommand{\eR}{\mathbb{R}}

\def\BibTeX{{\rm B\kern-.05em{\sc i\kern-.025em b}\kern-.08em
    T\kern-.1667em\lower.7ex\hbox{E}\kern-.125emX}}
\begin{document}

\title{R2D2 image reconstruction with model uncertainty quantification in radio astronomy
}
\author{\IEEEauthorblockN{
Amir Aghabiglou$^1$, Chung San Chu$^{1}$, Arwa Dabbech$^{1}$,
Yves Wiaux$^1$\IEEEauthorrefmark{2}\thanks{The work was supported by EPSRC under grants EP/T028270/1 and ST/W000970/1. Computing \mbox{resources} came from the Cirrus UK National Tier-2 HPC Service at EPCC (http://www.cirrus.ac.uk) funded by the University of Edinburgh and EPSRC (EP/P020267/1). 
}}
\IEEEauthorblockA{$^1$Institute of Sensors, Signals and Systems, Heriot-Watt University, Edinburgh EH14 4AS, United Kingdom 
}
\IEEEauthorblockA{Email: \IEEEauthorrefmark{2}y.wiaux@hw.ac.uk}
}

\maketitle

\begin{abstract}

The ``Residual-to-Residual DNN series for high-Dynamic range imaging'' (R2D2) approach was recently introduced for Radio-Interferometric (RI) imaging in astronomy. R2D2's reconstruction is formed as a series of residual images, iteratively estimated as outputs of Deep Neural Networks (DNNs) taking the previous iteration's image estimate and associated data residual as inputs. 
In this work, we investigate the robustness of the R2D2 image estimation process, by studying the uncertainty associated with its series of learned models. Adopting an ensemble averaging approach, multiple series can be trained, arising from different random DNN initializations of the training process at each iteration. The resulting multiple R2D2 instances can also be leveraged to generate ``R2D2 samples'', from which empirical mean and standard deviation endow the algorithm with a joint estimation and uncertainty quantification functionality. Focusing on RI imaging, and adopting a telescope-specific approach, multiple R2D2 instances were trained to encompass the most general observation setting of the Very Large Array (VLA). Simulations and real-data experiments confirm that: (i) R2D2's image estimation capability is superior to that of the state-of-the-art algorithms; (ii) its ultra-fast reconstruction capability (arising from series with only few DNNs) makes the computation of multiple reconstruction samples and of uncertainty maps practical even at large image dimension; (iii) it is characterized by a very low model uncertainty.
\end{abstract}

\begin{IEEEkeywords}
Computational methods , Deep learning, Astronomy image processing, Aperture synthesis
\end{IEEEkeywords}

\section{Introduction}
In the field of imaging sciences, one of the significant challenges is to address ill-posed imaging inverse problems while ensuring accurate reconstructions and scalability to large datasets. Aperture synthesis by RI in astronomy is an iconic example of this challenge towards joint precision and scalability. In RI, each pair of antennas captures noisy complex measurements, termed visibilities, corresponding to Fourier components of the sought radio emission \cite{thompson2017interferometry}. These measurements, gathered over the observation period, form an incomplete coverage of the 2D Fourier plane, constituting the Fourier sampling pattern.

Consider ${\xb^{\star}}\in\mathbb{R}_{+}^N$ as the unknown radio image of interest and focusing on monochromatic intensity imaging, 
% then
the observed data ${\yb}\in\eC^M$ can be expressed as:
\begin{equation}
{\yb}={\Phib} {\xb^{\star}}+\nb,
\label{eq:observation}
\end{equation}
where $\nb \in \eC^{M}$ represents complex random noise following a Gaussian distribution $\mathcal{N}(0,\tau^2)$, with $\tau>0$.
Here, ${\Phib}\colon \eR^N \to \eC^M$ is the measurement operator, describing the non-uniform Fourier sampling \cite{thompson2017interferometry} and often incorporating a data-weighting scheme to improve the observation's effective resolution \cite{briggs1995high}.

RI data can be formulated in the image domain via a back-projection, resulting in the \textit{dirty} image $\xb_{\textrm{d}} \in \eR^N$ given by:
\begin{equation}
\xb_{\textrm{d}} =\kappa \text{Re}\{\Phib^\dagger \yb\}=\kappa \text{Re}\{\Phib^\dagger\Phib \xb^{\star} + \Phib^\dagger \nb\},
\label{eq:backprojection}
\end{equation}
where $\kappa>0$ serves as a normalization factor, ensuring that the point spread function (PSF) has a peak value of one. The linear operator $\kappa \text{Re}\{\Phib^\dagger\Phib\} \colon \eR^N \to \eR^N $ maps the image of interest to the dirty image space, with $\text{Re}\{\cdot\}$ indicating the real part of this dirty image. The term $\kappa \text{Re}\{\Phib^\dagger \nb \}$ represents the image-domain noise vector.

The CLEAN algorithm \cite{hogbom1974aperture}, a bespoke version of the Matching Pursuit algorithm \cite{Mallat93} that iteratively identifies model components from residual dirty images, holds a near-total monopoly in the field. While its simplicity and computational efficiency endows the algorithm with a high degree of scalability, it also limits its capability in terms of reconstruction quality. Over the past decade, new RI imaging algorithms have emerged, aiming to supersede CLEAN. Optimization algorithms propelled by advanced sparsity-based regularization models have been proposed \cite{wiaux2009compressed,li2011application,carrillo2012sparsity,dabbech2015moresane,repetti2020forward}, with  uSARA the latest evolution \cite{terris2022}. 
Recent advances in deep learning have opened new avenues in RI imaging, offering both modeling power and acceleration. These range from end-to-end approaches such as POLISH \cite{connor2022deep, terris2022, Schmidt22}, to Plug-and-Play (PnP) algorithms such as AIRI \cite{terris2022}, whose regularization operators are encapsulated in DNN denoisers. 
While such optimization and PnP algorithms have demonstrated remarkable precision indeed, their highly iterative nature poses  challenges towards scaling to the extreme data volumes expected from modern telescopes such as the VLA\footnote{\url{https://public.nrao.edu/telescopes/vla/}} and its next-generation upgrade ngVLA.

In recent work, a novel deep learning-based algorithm was proposed for RI imaging, termed ``Residual-to-Residual DNN series for high-Dynamic range imaging'' (R2D2) \cite{aghabiglou2023b,aghabiglou24}. R2D2's reconstruction is formed as a series of residual images, iteratively estimated as outputs of regularization DNNs taking the previous iteration's image estimate and associated data residual as inputs. The method can be interpreted as a learned version of Matching Pursuit. In this work, we investigate the robustness of the R2D2 image estimation process,  studying the uncertainty associated with its series of learned models.

The remainder of this paper is structured as follows. In Section \ref{sec:method}, the methodology underlying the joint image and estimation and uncertainty quantification with R2D2 is summarized. Section \ref{sec:experiments} provides a the validation of the proposed methodology, focusing on RI imaging, and adopting a telescope-specific approach. We conclude in Section \ref{sec:conclusion}

\section{R2D2 methodology} \label{sec:method}
\subsection{Image Reconstruction}
At the core of R2D2 lies a series of DNNs $(\Nb_{\widehat{\thetab}^{(i)}})_{1 \leq i \leq I}$, each wielding the power of deep learning to enhance image estimates through learned parameters $\widehat{\thetab}^{(i)}\in \eR^{Q}$. At each iteration  $i \in \{1,\dots,I\}$, the network $\Nb_{\widehat{\thetab}^{(i)}}$ takes as input both the previous image estimate $\xb^{(i-1)}$ and its associated residual dirty image, given by: 
\begin{equation}
\rb^{(i-1)}=\xb_{\textrm{d}}-\kappa \text{Re}\{\Phib^\dagger\Phib\}\xb^{(i-1)}.
\label{eq:residual_update}
\end{equation}
% as input. 
At the first iteration, the initial image estimate and residual dirty image are initialized as
% is set to zero 
$\xb^{(0)}=0 \in \eR^{N}$ and $\rb^{(0)}=\xb_{\textrm{d}}$. Subsequently, the image estimate is updated by adding the output of the network to the previous estimate:
\begin{equation}
\xb^{(i)}=\xb^{(i-1)} + {\Nb_{\widehat{\thetab}^{(i)}}}(\rb^{(i-1)}, \xb^{(i-1)}).
\label{eq:image_update}
\end{equation}
 
This iterative refinement gradually improves the resolution and dynamic range of the reconstruction. Consequently, the final reconstructed image $\widehat{\xb}$ is a ``series'' formed by the summation of output residual images from all network components.  

Training a sequence of R2D2 DNNs $({\Nb_{\widehat{\thetab}^{(i)}}})_{1 \leq i \leq I}$ commences with a dataset comprising $K$ ground truth images ${\xb}^{\star}_{k}$, from which simulated measurements $({\yb}_{k})_{1 \leq i \leq K}$ are generated following the procedure outlined in \eqref{eq:observation}. Subsequently, at each $1 \leq i \leq I$, a set of residuals ${\rb}^{(i-1)}_{k}$ is created via \eqref{eq:residual_update}.
% , and then train ${\Nb_{\widehat{\thetab}^{(i)}}}$ to 
Each DNN is trained by minimizing the loss function defined as:
\begin{equation}
\label{eq:r2d2seriesloss}
  \min_{{\thetab}^{(i)}\in \eR^Q}\frac{1}{K} \sum_{k=1}^{K} ~ \lVert  {\xb}^{\star}_{k} - [{\xb}^{(i-1)}_{k} + {{{\Nb}}_{{\thetab}^{(i)}}(\rb_{k}^{(i-1)},{\xb}^{(i-1)}_{k})}]_{+} \rVert_{1},
\end{equation}
where $\|.\|_1$ denotes the $\ell_1$-norm. To maintain the positivity of the image estimate during each iteration, the operator $[.]_{+}$ projects the image estimate onto the positive orthant $\mathbb{R}^N_+$.
% , denoted by $[.]_{+}$, is utilized.

\subsection{Robustness \& Uncertainty Quantification}

Uncertainty quantification is of paramount importance in ill-posed inverse imaging problems. On the one hand, the incompleteness of the observed data unavoidably leads to uncertainty, aka aleatoric uncertainty. 
It can typically be studied by sampling from the posterior distribution $p(\xb | \yb)$, derived from \eqref{eq:observation} using Bayes' theorem, with a sampling algorithm, such as Monte Carlo Markov Chain (MCMC) methods \cite{laumont2022bayesian,mukherjee2023learned}. The uncertainty can then be described by the statistics of the samples, such as their mean and standard deviation. On the other hand, the uncertainty associated with the choice of a specific regularization model, aka epistemic uncertainty, is no less important.

The R2D2 algorithm is deterministic in nature, in which context the quantification of aleatoric uncertainty is not directly accessible. We instead propose in this work to investigate the epistemic uncertainty in the reconstruction, associated with the choice regularization model encapsulated in the DNNs. Adopting an ensemble averaging approach, multiple series can be trained, arising from different random DNN initializations of the training process at each iteration. The resulting multiple R2D2 instances can also be leveraged to generate ``R2D2 samples'', from which empirical mean and standard deviation endow the algorithm with a joint estimation and uncertainty quantification functionality. 
Similar epistemic uncertainty quantification approaches have been proposed in
\cite{terris2023plug,narnhofer2021bayesian,lahlou2021deup}. 

\section{Validation} \label{sec:experiments}

\subsection{Training Dataset and Training}
As in \cite{aghabiglou24}, our ground truth dataset comprises 20000 images of size $N = 512 \times 512$ and of peak pixel value $1$, curated from astronomical images at %both optical and radio 
optical wavelengths, as well as medical images, pre-processed following \cite{terris2023plug}. We define the dynamic range $a$ of an image of interest as the ratio of its peak to the faintest pixel intensity, or similarly the standard deviation of the background noise. In order to ensure that the range of dynamic ranges present in the ground truth dataset matches the extremely high values expected in RI, typically $a \in [10^3, 5\times 10^5]$, pixel-wise exponentiation procedures are applied, as described in \cite{terris2022}. Adopting the same VLA-specific approach undertaken in \cite{aghabiglou24}, we sample a different Fourier sampling pattern per each ground truth image, resulting in 20000 dirty images as per \eqref{eq:backprojection}. The measurement noise is adapted to ensure that the back-projected noise level is commensurate to the inverse dynamic range of the exponentiated ground truth image. A comprehensive depiction of the training procedure, including data fidelity-driven pruning of the training dataset, is outlined in \cite{aghabiglou24}. Our core network architecture for each DNN is the popular U-Net DNN \cite{Ronneberger15}, optimized for each R2D2 iteration using \eqref{eq:r2d2seriesloss}. Specific to the target uncertainty quantification analysis, and targeting a simple proof of concept, we train $L=5$ independent realizations of the R2D2 DNN series.

\subsection{Simulations: Test Dataset}  
The test dataset includes four authentic radio images, namely 3C353 (sourced from NRAO Archives), Messier 106 \cite{shimwell2022}, Abell 2034 and PSZ2 G165.68+44.01 \cite{botteon2022}, from where 200 ground truth images of size $N = 512 \times 512$ were generated  following the same pre-processing and exponentiation steps employed for the training dataset. Corresponding dirty images $\xb_{\textrm{d}}$ were created by applying an unique sampling pattern to each ground truth image via \eqref{eq:backprojection}. 

\subsection{Simulations: Benchmark algorithms} 

The joint R2D2 functionality for image estimation and uncertainty quantification will be validated, in comparison to: (i) a U-Net trained end-to-end to deliver a reconstruction directly from the dirty image (which is nothing else than R2D2's first iteration); (ii) uSARA, AIRI, and CLEAN (delivering image estimation only).

\subsection{Simulations: Metrics} 

We assess image reconstruction quality using the signal-to-noise ratio (SNR) metric, given by:
\begin{equation}
\textrm{SNR}(\widehat{\xb}, {\xb^{\star}}) = 20\textrm{log}_{10}(\| {\xb^{\star}}\|_2 / \|{\xb^{\star}} - \widehat{\xb}\|_2), 
\end{equation}
where $\widehat{\xb}$ is the reconstruction by any algorithm with ground truth $\xb^{\star}$. 
To examine the performance of an algorithm in high-dynamic range scenario and to facilitate visualization of faint features, we transform an image $\xb$ to logarithmic scale by:
\begin{equation}
\operatorname{rlog}(\xb) = x_{\textrm{max}}\textrm{log}_{a}(\frac{a}{x_{\textrm{max}}}~ \xb+\textbf{1}),
\end{equation}
where $a$ is the dynamic range of the image of interest and $x_{\textrm{max}}$ is the peak pixel value of $\xb$. The SNR metric in logarithmic scale is given by:
\begin{equation}
\textrm{log}\textrm{SNR}(\widehat{\xb}, {\xb^{\star}}) = \textrm{SNR}(\textrm{rlog}(\widehat{\xb}), \textrm{rlog}({\xb^{\star}})).
\end{equation}

The set of image reconstructions from the $L=5$ realizations of R2D2 read:
\begin{equation}
    \bar{\xb}^{(i)} = [\hat{\xb}_1^{(i)}, \dots, \hat{\xb}_L^{(i)}],
\end{equation}
where $\hat{\xb}_\ell^{(i)}$ is the estimate from the $\ell$-th R2D2 realization at the $i$-th iteration. To investigate the associated uncertainty, we evaluate the pixel-wise mean $\mu(\bar{\xb})$, given by:
\begin{equation}
    \mu(\bar{\xb}^{(i)}) = \frac{1}{L}\sum_{\ell=1}^L \hat{\xb}_{\ell}^{(i)},
\end{equation}
as well as the pixel-wise ratio of the standard deviation to mean, denoted as $[\sigma/\mu](\bar{\xb})$, which reads:
\begin{equation}
    [\sigma/\mu](\bar{\xb}^{(i)}) = \begin{cases}
        \frac{1}{\mu(\bar{\xb}^{(i)})}\sqrt{\frac{\sum_1^L(\xb_{\ell}^{(i)} - \mu(\bar{\xb}^{(i)}))^2}{L}} & \textrm{if } \mu(\bar{\xb}^{(i)}) < 1/a, \\
        0 & \textrm{otherwise}.
    \end{cases}
\end{equation}

\subsection{Simulations: Results \& Discussion}
\begin{figure}[]
\centering
\setlength\tabcolsep{1pt}
\begin{tabular}{ccc}
    \includegraphics[width=0.33\linewidth]{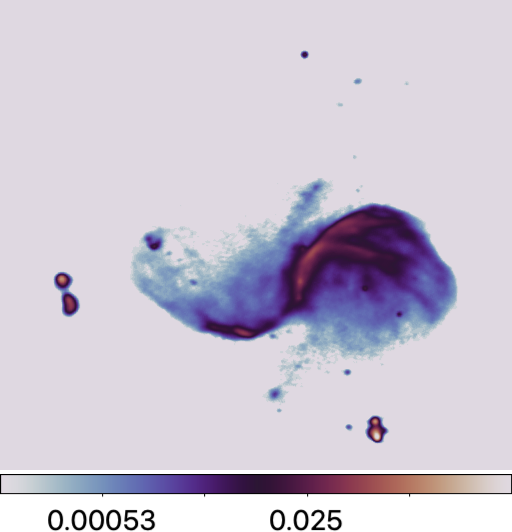}&
    \includegraphics[width=0.33\linewidth]{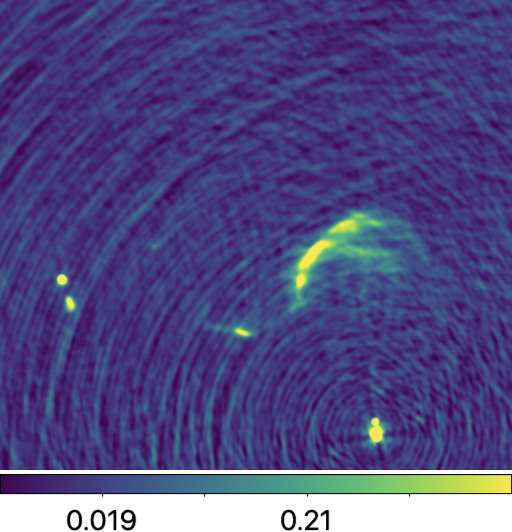}&
    \includegraphics[width=0.33\linewidth]{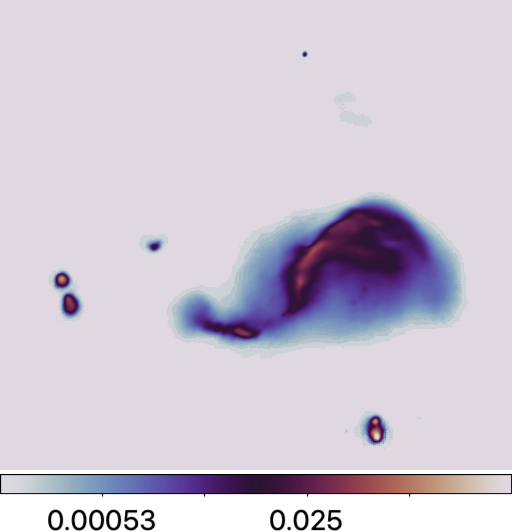}\\
    Ground truth $\xb^{\star}$ & ${\rb_{\textrm{R2D2}}^{(0)}}$ & ${\hat{\xb}^{(1)}}_{\textrm{R2D2}}$\\
     \includegraphics[width=0.33\linewidth]{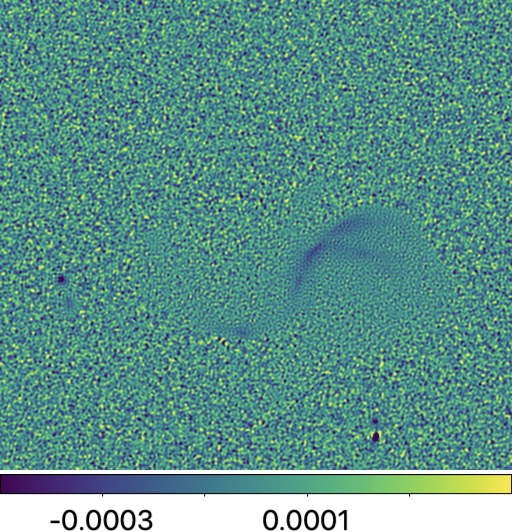}&
    \includegraphics[width=0.33\linewidth]{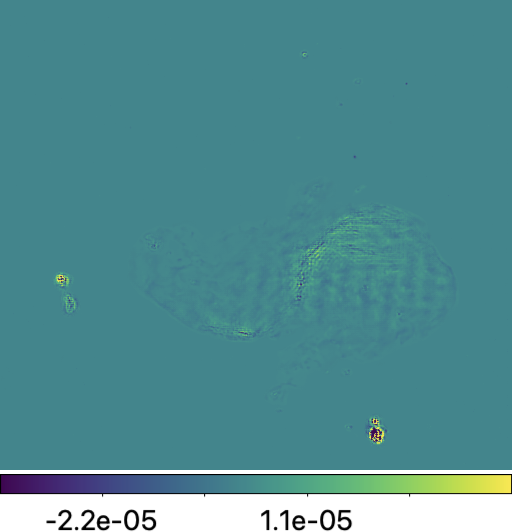}&
    \includegraphics[width=0.33\linewidth]{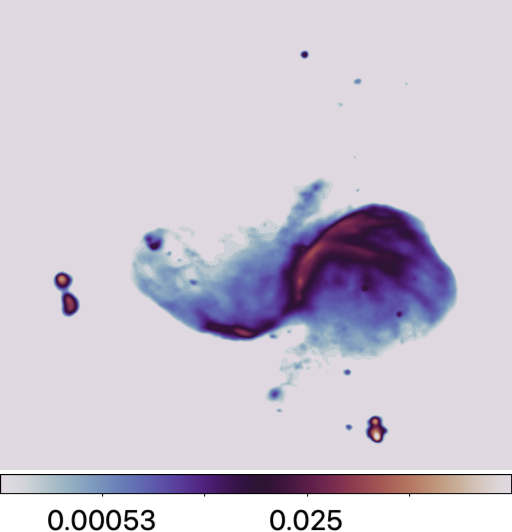}\\
    $\rb^{(11)}_{\textrm{R2D2}}$& $\hat{\xb}_{\textrm{R2D2}}^{(12)} -\hat{\xb}_{\textrm{R2D2}}^{(11)}$ %${{\Nb}}_{\widehat{\thetab}^{(12)}}(\rb^{(11)},\hat{\xb}^{(11)})$ 
    & ${\hat{\xb}^{(12)}}_{\textrm{R2D2}}$
    \\ 
\end{tabular}
\caption{Evolution of the R2D2 series (single realization) across its iterations. The ground truth image $\xb^{\star}$ and dirty image $\rb^{(0)}_{\textrm{R2D2}}$  are reported in the top left and center panels, respectively. The output of the first DNN gives $\hat{\xb}^{(1)}_{\textrm{R2D2}}$ (top right). 12 iterations later, $\rb^{(11)}_{\textrm{R2D2}}$ (bottom left) and $\hat{\xb}^{(12)}_{\textrm{R2D2}}$ are fed to the last DNN, whose output (bottom center) is added to $\hat{\xb}^{(12)}_{\textrm{R2D2}}$ to deliver the reconstruction (bottom right).}
    \label{fig:r2d2_evolve}
\end{figure}

Fig.~\ref{fig:r2d2_evolve} illustrates the progression of the R2D2 series from its initial to final iteration, simultaneously peeling residuals $\rb^{(i)}_{\textrm{R2D2}}$ and updating the image estimate $\hat{\xb}^{(i)}_{\textrm{R2D2}}$ from the DNN output. 

\begin{figure*}
    \centering
    \setlength\tabcolsep{1pt}
    \begin{tabular}{ccc}
        \includegraphics[width=0.33\linewidth]{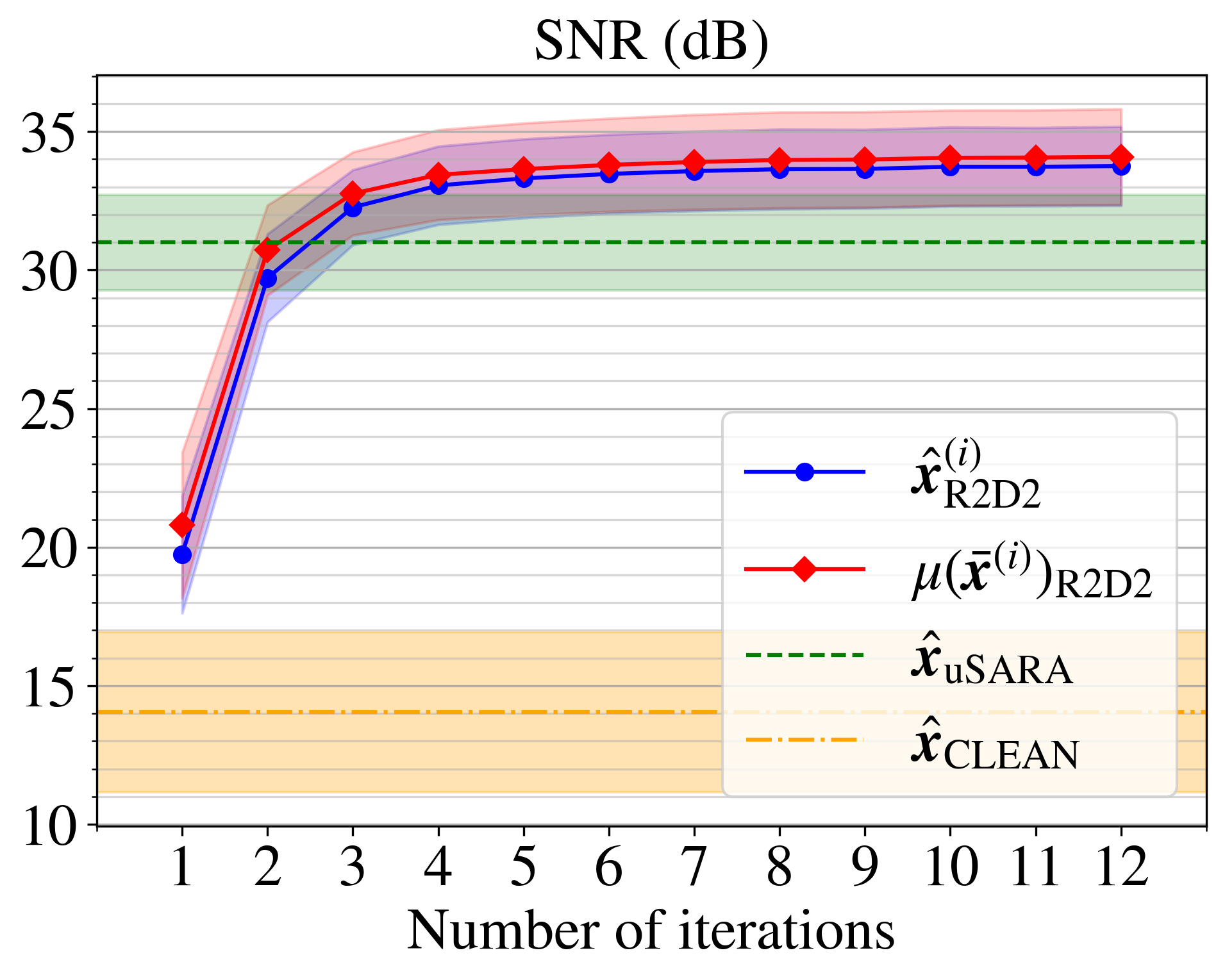} & 
        \includegraphics[width=0.33\linewidth]{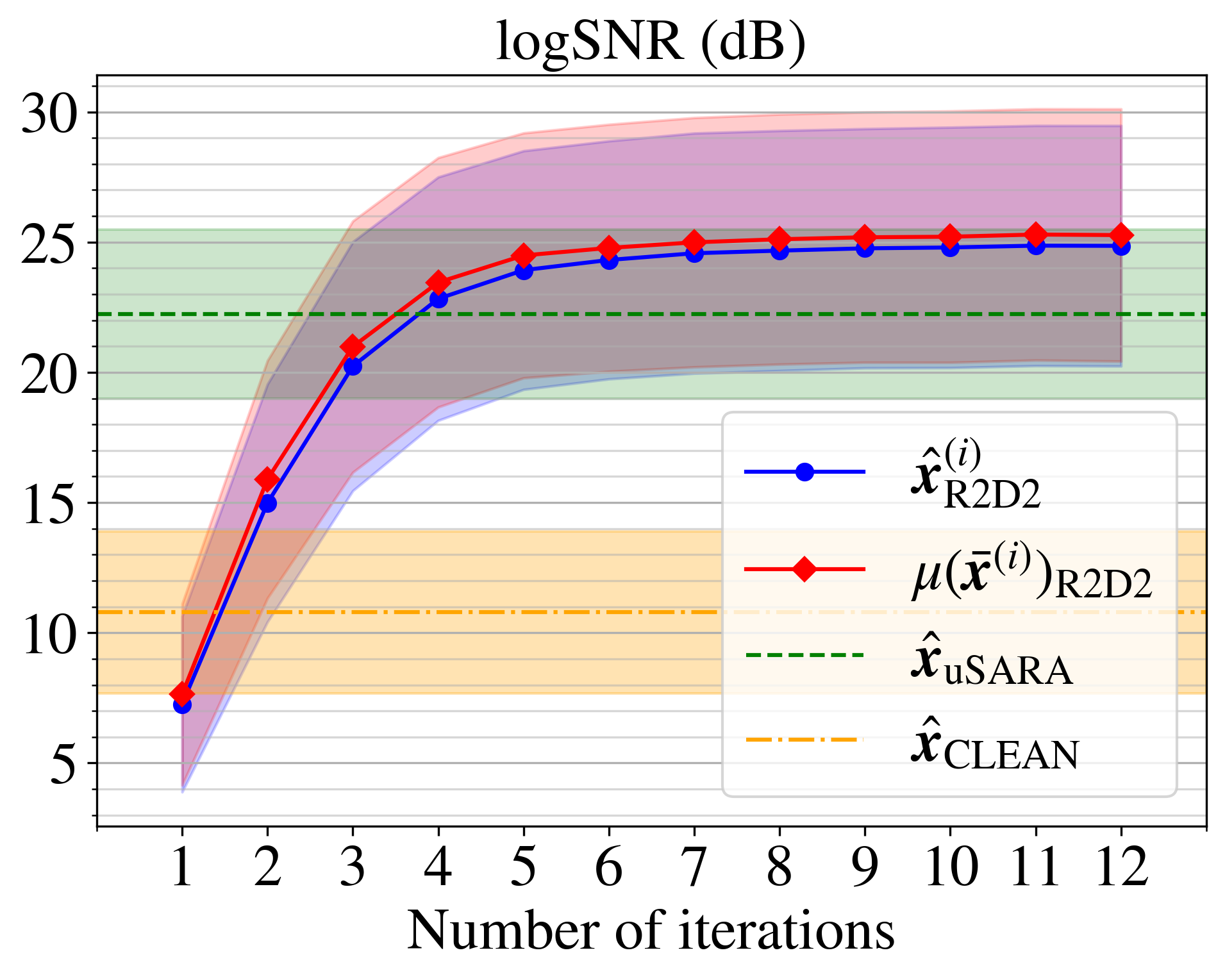} & 
        \includegraphics[width=0.33\linewidth]{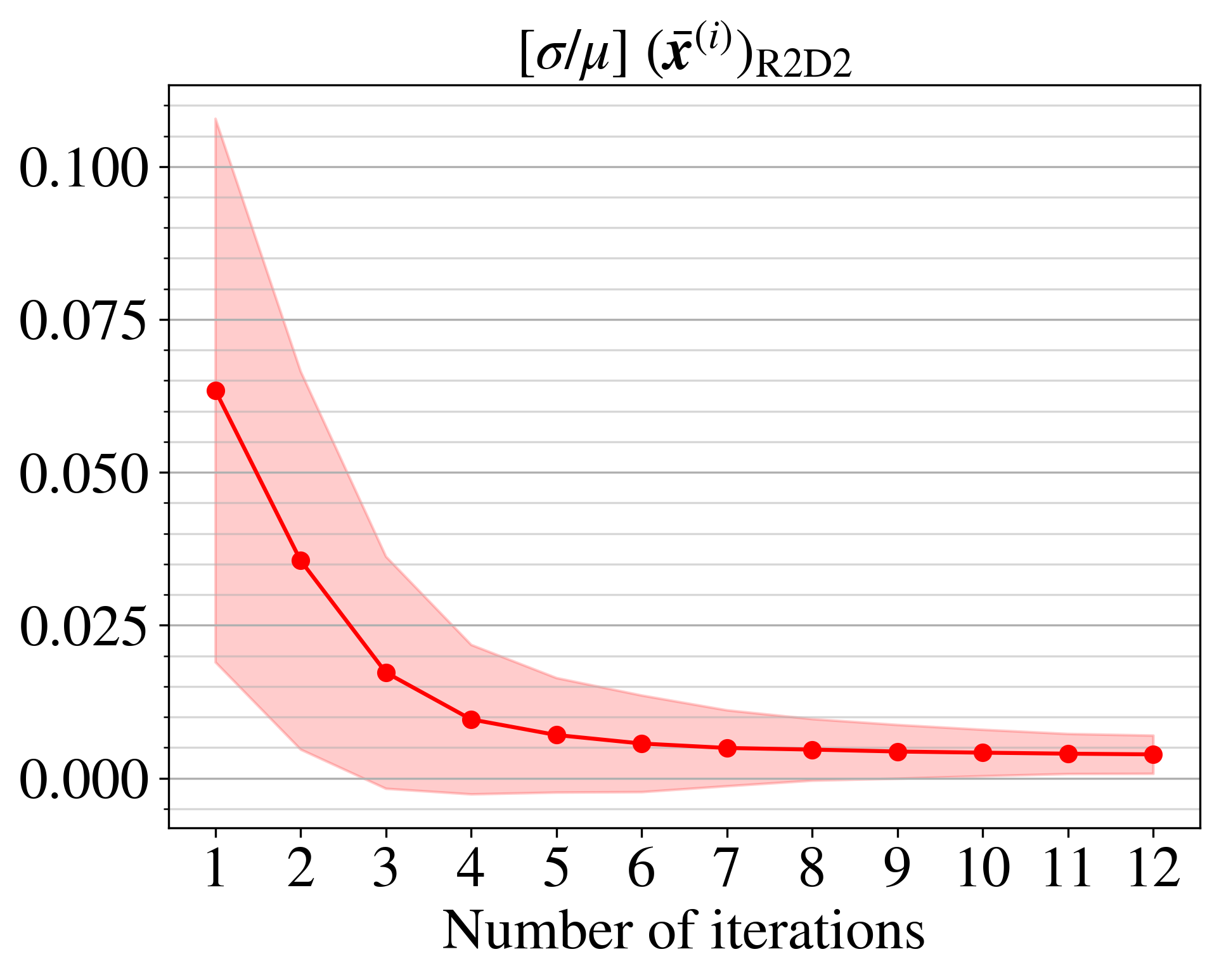}
    \end{tabular}
    \caption{Quantitative metric analysis. Left (resp.~center): Progress of the reconstruction quality evaluated by SNR (resp.~logSNR) across R2D2 iterations for both $\mu(\bar{\xb}^{(i)})_{\textrm{R2D2}}$ and ${\hat{\xb}^{(i)}}_{\textrm{R2D2}}$. Right: Progress of the epistemic uncertainty evaluated  by $[\sigma/\mu](\bar{\xb}^{(i)})_{\textrm{R2D2}}$ across iterations. Values of the SNR and logSNR, achieved at convergence by the benchmark algorithms uSARA and CLEAN, are reported via horizontal lines, in their respective plots. Each point is the averaged metric over 200 inverse problems (5$\times$200 for $\hat{\xb}_{\textrm{R2D2}}$), while the shaded area presents its standard deviations.}
    \label{fig:SNRplots}
\end{figure*}

The left and center panels of Fig.~\ref{fig:SNRplots} present the graphs of the SNR and logSNR metrics across iterations, respectively. For R2D2, these metrics are computed both for the image estimate ${\hat{\xb}^{(i)}}_{\textrm{R2D2}}$ of a single R2D2 realization and for their pixel-wise mean $\mu(\bar{\xb}^{(i)})_{\textrm{R2D2}}$. 
The U-Net results are those of the first R2D2 iteration. Final results of uSARA and CLEAN, respectively denoted $\hat{\xb}_{\textrm{uSARA}}$ and $\hat{\xb}_{\textrm{CLEAN}}$, are shown as horizontal lines. For simplicity, AIRI results, on par with those of uSARA, are not reported. The mean and standard deviation across the 200 test inverse problems and 5 R2D2 realizations are reported for $\mu(\bar{\xb}^{(i)})_{\textrm{R2D2}}$, and 
simply across the 200 test inverse problems for ${\hat{\xb}^{(i)}}_{\textrm{R2D2}}$, $\hat{\xb}_{\textrm{uSARA}}$, and $\hat{\xb}_{\textrm{CLEAN}}$. Firstly, the results confirm that R2D2's image estimation capability is superior to that of the state-of-the-art algorithms. Secondly, in both SNR and logSNR, $\mu(\bar{\xb}^{(i)})_{\textrm{R2D2}}$ yields slightly superior results compared to ${\hat{\xb}^{(i)}}_{\textrm{R2D2}}$, suggesting an advantage in averaging ``R2D2 samples'', in terms of precision.

Table~\ref{tab:timing} reports the computational cost for different imaging algorithms, in terms of number of iterations required and total reconstruction time. 
In addition to the higher SNR and logSNR values shown in Fig.~\ref{fig:SNRplots} when compared to the benchmarks, one can note that R2D2 also offers significant computational efficiency, requiring drastically fewer iterations than uSARA and AIRI, and taking only a fraction of time required by CLEAN, thanks to its ultra-fast DNN-encapsulated regularization. 
We note that U-Net and R2D2 are executed on a single GPU, 
CLEAN and uSARA are executed on a single CPU, while AIRI is implemented with  regularization DNNs on GPU and data-fidelity steps on CPU. These results confirm that R2D2 boasts ultra-fast reconstruction capability, which makes the computation of ``R2D2 samples'' practical even at large image dimension.

\begin{table}
    \centering
    \resizebox{0.8\columnwidth}{!}{\begin{tabular}{ccc}
        \toprule
        Algorithm & $I \pm $ std & $t_{\textrm{tot.}} \pm$ std (sec) \\
        \midrule
        $\hat{\xb}_\textrm{U-Net}$ & 1 & 0.1 $\pm$ 0.1  \\
        $\hat{\xb}_\textrm{R2D2}$ & 12 & 2.0 $\pm$ 0.4 \\
        $\hat{\xb}_\textrm{AIRI}$ & 4995$\pm$50 & 3430$\pm$1461 \\
        $\hat{\xb}_\textrm{uSARA}$ & 1107$\pm$377 & 4015.9$\pm$1471.2 \\
        $\hat{\xb}_\textrm{CLEAN}$ & 8$\pm$1& 93.2$\pm$ 26.7\\
        \bottomrule
    \end{tabular}}
    \caption{%Imaging 
    Computational cost of each imaging algorithm used in this study, in terms of number of iterations $I$ and total reconstruction time $t_{\textrm{tot.}}$ in seconds.}
    \label{tab:timing}
\end{table}

The graph of the relative pixel-wise epistemic uncertainty $[\sigma/\mu](\bar{\xb}^{(i)})_{\textrm{R2D2}}$, more specifically its mean and standard deviation across the 200 test inverse problems, is presented in the right panel of Fig.~\ref{fig:SNRplots}. A consistent decrease of $[\sigma/\mu](\bar{\xb}^{(i)})_{\textrm{R2D2}}$ across the R2D2 iterations is observed. This observation serves as compelling evidence that the R2D2 robustness (resp.~epistemic uncertainty) increases (resp.~decreases) rapidly with the number of terms in its series. 

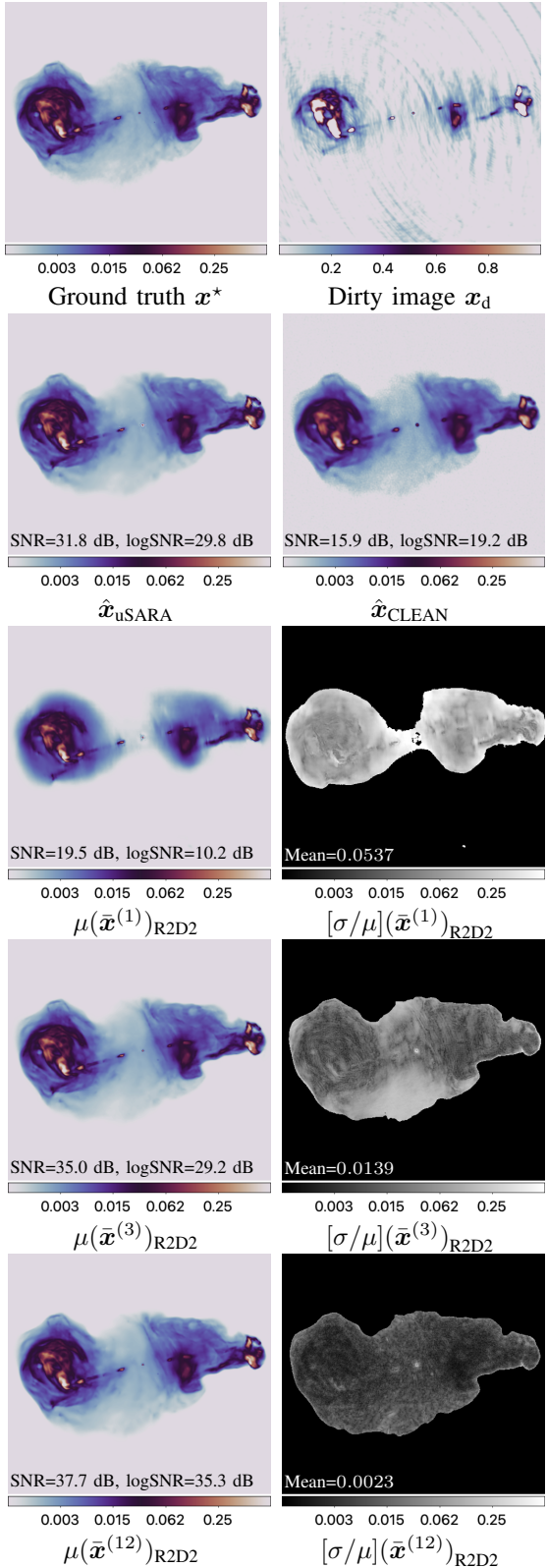
\begin{figure}
    \centering
    \setlength\tabcolsep{1pt}
    \begin{tabular}{cc}
        \includegraphics[width=0.4\linewidth]{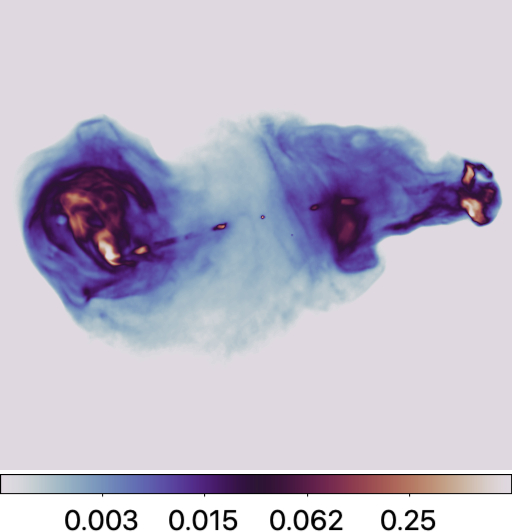} & 
        \includegraphics[width=0.4\linewidth]{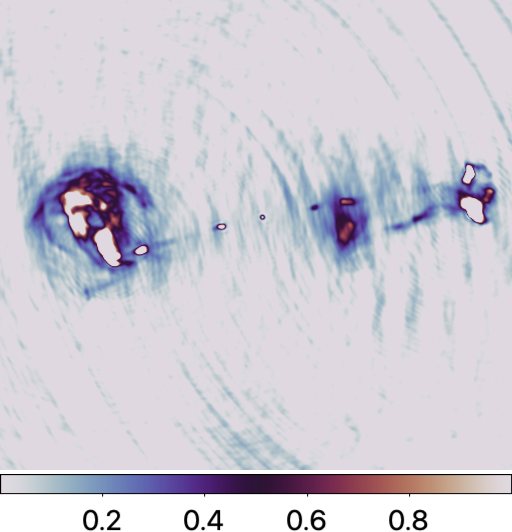} \\
        Ground truth ${\xb}^{\star}$ & Dirty image ${\xb}_{\textrm{d}}$ \\
        \begin{tikzpicture}
            \node[inner sep=0pt] at (0,0) {\includegraphics[width=0.4\linewidth]{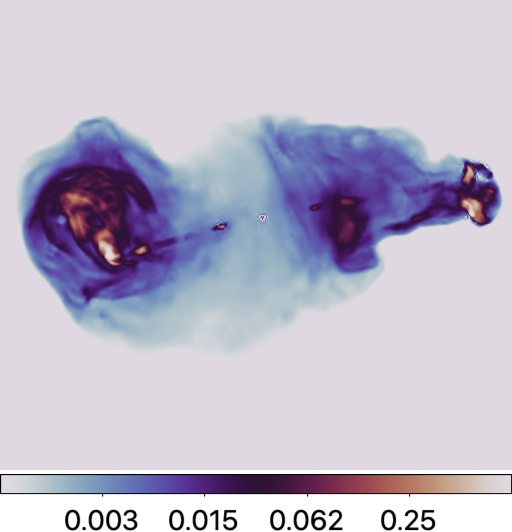}};
            \node at (-0.1,-1.25) {\textcolor{black}{\scriptsize{SNR=31.8 dB, logSNR=29.8 dB}}};
        \end{tikzpicture} &
        \begin{tikzpicture}
            \node[inner sep=0pt] at (0,0) {\includegraphics[width=0.4\linewidth]{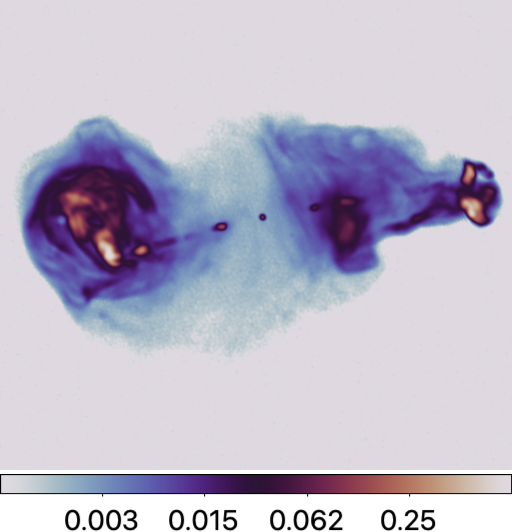}};
            \node at (-0.1,-1.25) {\textcolor{black}{\scriptsize{SNR=15.9 dB, logSNR=19.2 dB}}};
        \end{tikzpicture}\\
        $\hat{\xb}_{\textrm{uSARA}}$& $\hat{\xb}_{\textrm{CLEAN}}$\\
        % (31.8, 29.8)&(15.9, 19.2) \\
        % \includegraphics[width=0.48\linewidth]{fig/U-Net_3c353_avg.png}
        \begin{tikzpicture}
            \node[inner sep=0pt] at (0,0) {\includegraphics[width=0.4\linewidth]{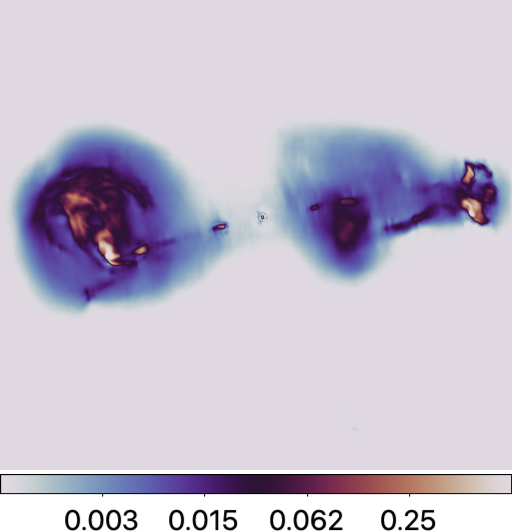}};
            \node at (-0.1,-1.25) {\textcolor{black}{\scriptsize{SNR=19.5 dB, logSNR=10.2 dB}}};
        \end{tikzpicture}& 
        \begin{tikzpicture}
            \node[inner sep=0pt] at (0,0) {\includegraphics[width=0.4\linewidth]{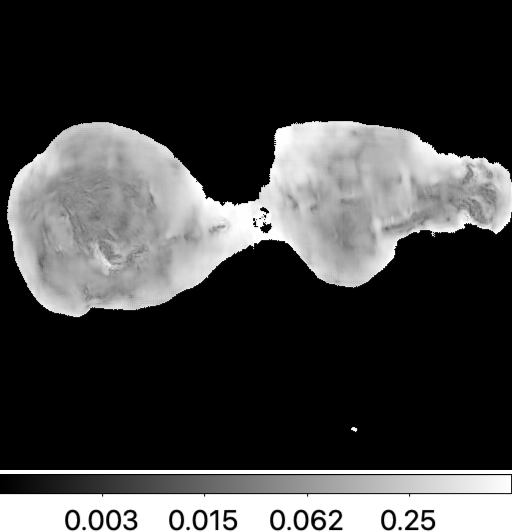}};
            \node at (-1,-1.25) {\textcolor{white}{\scriptsize{Mean=$0.0537$}}};
        \end{tikzpicture} \\
        ${\mu(\bar{\xb}^{(1)}})_{\textrm{R2D2}}$ & ${[\sigma/\mu](\bar{\xb}^{(1)})}_{\textrm{R2D2}}$ \\
        \begin{tikzpicture}
            \node[inner sep=0pt] at (0,0) {\includegraphics[width=0.4\linewidth]{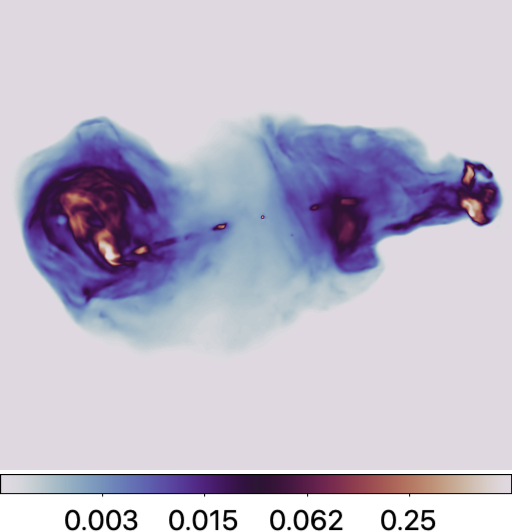}};
            \node at (-0.1,-1.25) {\textcolor{black}{\scriptsize{SNR=35.0 dB, logSNR=29.2 dB}}};
        \end{tikzpicture}& 
        \begin{tikzpicture}
            \node[inner sep=0pt] at (0,0) {\includegraphics[width=0.4\linewidth]{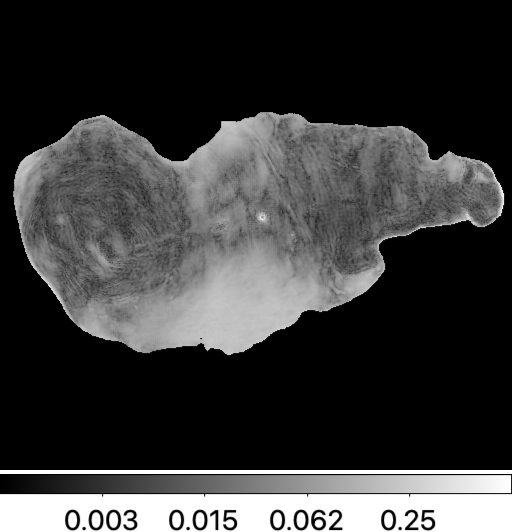}};
            \node at (-1,-1.25) {\textcolor{white}{\scriptsize{Mean=$0.0139$}}};
        \end{tikzpicture} \\
        ${\mu(\bar{\xb}^{(3)}})_{\textrm{R2D2}}$ & ${[\sigma/\mu](\bar{\xb}^{(3)})}_{\textrm{R2D2}}$\\
        
        \begin{tikzpicture}
            \node[inner sep=0pt] at (0,0) {\includegraphics[width=0.4\linewidth]{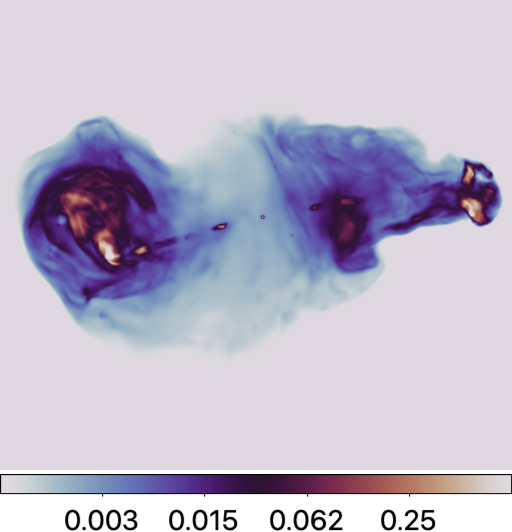}};
            \node at (-0.1,-1.25) {\textcolor{black}{\scriptsize{SNR=37.7 dB, logSNR=35.3 dB}}};
        \end{tikzpicture} &
        \begin{tikzpicture}
            \node[inner sep=0pt] at (0,0) {\includegraphics[width=0.4\linewidth]{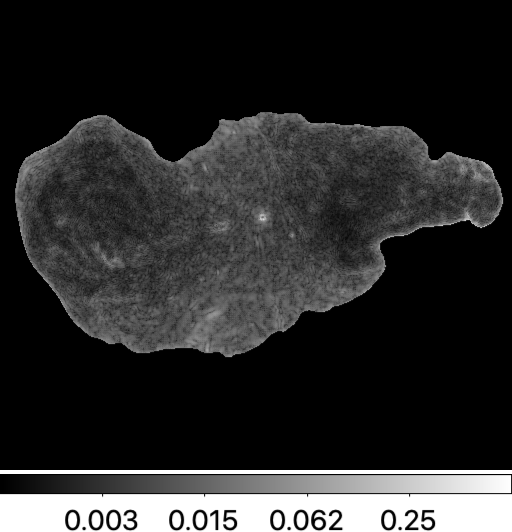}};
            \node at (-1,-1.25) {\textcolor{white}{\scriptsize{Mean=$0.0023$}}};
        \end{tikzpicture}\\
        
        ${\mu(\bar{\xb}^{(12)}})_{\textrm{R2D2}}$ \vspace{0.1em} &
         ${[\sigma/\mu](\bar{\xb}^{(12)})}_{\textrm{R2D2}}$ \\
    \end{tabular}
    \caption{Visual illustration of the R2D2 joint image estimation and uncertainty quantification functionality across iterations for the simulation results. %Sensitivity of imaging results to the training initialization of R2D2 algorithm.
    Top row: ground truth (left) and dirty image (right). Second row: reconstruction $\hat{\xb}_{\textrm{uSARA}}$ (left) and $\hat{\xb}_{\textrm{CLEAN}}$ (right). Third (resp.~fourth, fifth) row: $\mu(\bar{\xb}^{(i)})_{\textrm{R2D2}}$ (left) and $[\sigma/\mu](\bar{\xb}^{(i)})_{\textrm{R2D2}}$ (right) for iterations $i=1$ (U-Net) (resp.~$i=3$, $i=12$). The evaluation metrics (SNR, logSNR) are embedded inside each image estimate. The mean pixel values across the image is shown inside each uncertainty image.}
    \label{fig:result_sim_3c353}
\end{figure}

In Fig.~\ref{fig:result_sim_3c353}, estimated images are presented for the various algorithms under scrutiny, and associated relative uncertainty maps $[\sigma/\mu](\bar{\xb}^{(i)})_{\textrm{R2D2}}$ for the R2D2 model, for one of the 200 inverse problems in the test dataset.
% same set of 200 inverse problems utilized in Figs. \ref{fig:overall} and \ref{fig:pixelstdovermean}. 
The first row displays the ground truth ${\xb}^{\star}$ (left) and the dirty image ${\xb}_{\textrm{d}}$ (right), the second row $\hat{\xb}_{\textrm{uSARA}}$ (left), and $\hat{\xb}_{\textrm{CLEAN}}$ (right). The third (resp.~forth, fifth) row displays $\mu(\bar{\xb}^{(i)})_{\textrm{R2D2}}$ (left) and $[\sigma/\mu](\bar{\xb}^{(i)})_{\textrm{R2D2}}$ (right) at iterations $i=1$ (U-Net) (resp.~$i=3$, $i=12$). AIRI results, on par with those of uSARA, are not displayed. For the R2D2 uncertainty maps, the mean pixel value is reported on the image itself, highlighting a relative uncertainty of the order 0.2\% only for R2D2, against 5\% for U-Net. These results corroborate the analysis of the metrics in Fig.~\ref{fig:SNRplots}. 

\subsection{Real Data Proof of Concept} \label{sec:real_data}
\begin{figure}
\setlength\tabcolsep{1pt}
    \centering
    \begin{tabular}{cc}
        \includegraphics[width=0.48\linewidth]{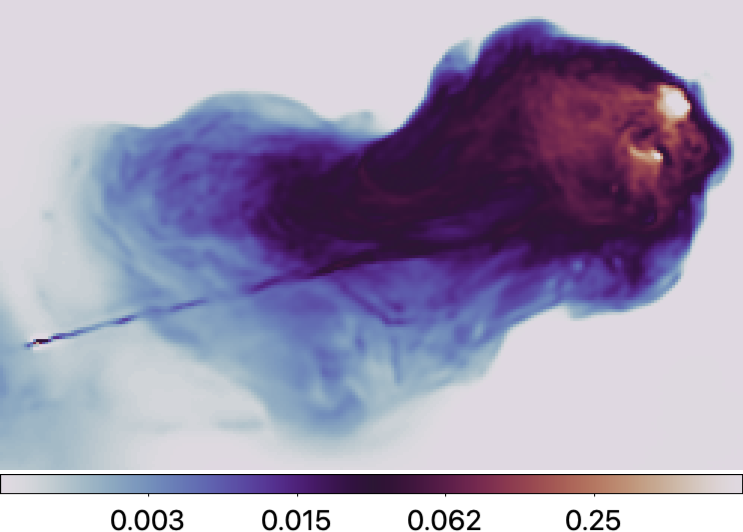} &
        \begin{tikzpicture}
            \node[inner sep=0pt] at (0,0) {\includegraphics[width=0.48\linewidth]{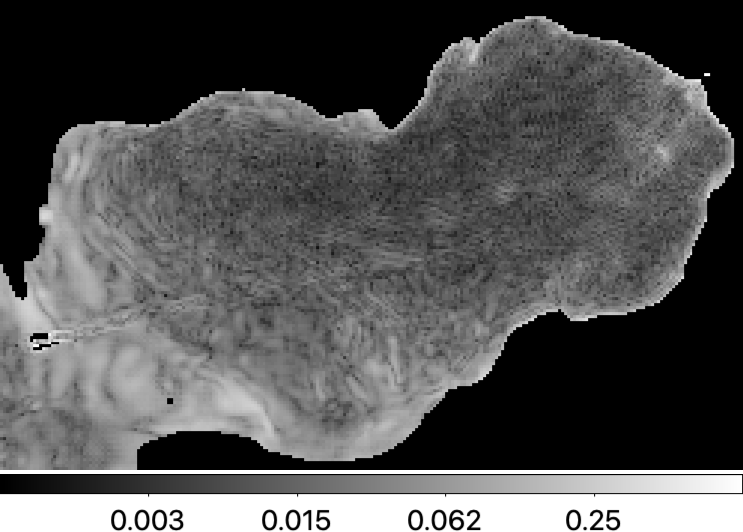}};
            \node at (-1.4,1.3) {\textcolor{white}{\scriptsize{Mean=$0.005$}}};
        \end{tikzpicture}\\
        $\mu(\bar{\xb}^{(12)})_{\textrm{R2D2}}$ & 
        $[\sigma/\mu](\bar{\xb}^{(12)})_{\textrm{R2D2}}$\\
    \end{tabular}
    \caption{Visual illustration of the R2D2 joint image estimation and uncertainty quantification functionality across iterations for the real data results. Left: pixel-wise mean $\mu(\bar{\xb}^{(12)})_{\textrm{R2D2}}$. Right: uncertainty map $[\sigma/\mu](\bar{\xb}^{(12)})_{\textrm{R2D2}}$.}
    \label{fig:real_data}
\end{figure}

In this section, the joint image estimation and uncertainty quantification functionality of R2D2 is validated on real data. R2D2 is applied to image of the radio galaxy Cygnus~A from S band observations with the Very Large Array (VLA; see \cite{aghabiglou2023b} for detailed data description). The left panel of Fig.~\ref{fig:real_data} displays the pixel-wise mean image $\mu(\bar{\xb}^{(12)})_{\textrm{R2D2}}$ (zoom on the right lobe of the galaxy) derived from 5 reconstructions generated by the R2D2 algorithm. The reconstruction exhibits high resolution across both low and high intensity values, illustrating the algorithm's precision. The right panel displays the associated uncertainty map $[\sigma/\mu](\bar{\xb}^{(12)})_{\textrm{R2D2}}$ (zoom on the same region). The mean pixel value is reported in the upper left corner, highlighting a relative uncertainty of the order 0.5\% only.

\section{Conclusion} \label{sec:conclusion}

In this work, we have investigated the robustness of the R2D2 image estimation process, adopting an ensemble averaging approach and studying the uncertainty arising from the random initialization of the training process for each DNN of the series. Focusing on high-dynamic range RI imaging in astronomy, 5 R2D2 series realizations with 12 terms each  were trained for the VLA. Simulations and real-data experiments confirm that: (i) R2D2's image estimation capability is largely superior to that of CLEAN, U-Net, and even uSARA and AIRI; (ii) R2D2 boasts ultra-fast reconstruction capability, making the computation of multiple reconstruction samples and of uncertainty maps practical even at large image dimension; (iii) it is characterized by a very low epistemic uncertainty (contrasting to the first iteration of the series, \emph{i.e.}~U-Net).

\bibliographystyle{IEEEtran}
\bibliography{main}

\end{document}